# The Critical Role of Thermal Fluctuations for Electrocatalytic Metal Surface Properties and CO Binding Trends

Wan-Lu Li,[1-3] Christianna N. Lininger,[1,2,4] Valerie Vaissier Welborn,[1-3,6] Elliot Rossomme,[2,3] Alexis T. Bell,[1,4] Martin Head-Gordon,[1-3] Teresa Head-Gordon[1-5*]

[1]Chemical Sciences Division, Lawrence Berkeley National Laboratory, Berkeley, California 94720, USA
[2]Kenneth S. Pitzer Center for Theoretical Chemistry, University of California, Berkeley, California 94720, USA
[3]Department of Chemistry, [4]Department of Chemical and Biomolecular Engineering, and [5]Department of Bioengineering, University of California, Berkeley, California 94720, USA
[6]Department of Chemistry, Virginia Tech, Blacksburg, VA 26067, USA

## Abstract

This work addresses a longstanding theoretical discrepancy using Density Functional Theory (DFT) with experimental observations of CO binding trends on electrocatalytically relevant metals for the $CO_2$ reduction reaction (CO2RR). By introducing thermal fluctuations using appropriate statistical mechanical NVT and NPT ensembles, we show that DFT with universal dispersion interactions yields qualitatively better metal surface strain trends and CO binding energetics, consistently predicts the correct site preference for all metals due to thermally induced surface distortions that preferentially exposes the undercoordinated atop site for Cu(111) and Pt(111), and for the weak binding Ag(111) and Au(111) surfaces at finite temperatures shows CO-metal interactions that are a mixture of chemisorbed and physisorbed species. This study better places theory as an equal partner to experimental heterogeneous catalysis by demonstrating the need to fully account for finite temperature fluctuations to make contact with surface science experiments.

**Corresponding Author**
*thg@berkeley.edu

# INTRODUCTION

The electrochemical CO$_2$ reduction reaction (CO2RR) is one of the most promising technologies available for converting greenhouse gases into useful chemicals using renewable energy sources.(*1, 2*) Given the low concentration of CO$_2$ adsorbed at the active metal interface, the inherent mechanism of CO2RR has proved to be difficult to investigate experimentally, although new spectroscopic surface sensitive and *operando* measurements are starting to emerge(*3*). It has also been a long–standing theoretical challenge to accurately predict the CO2RR reaction mechanism by faithfully reproducing the binding energies and activation barriers observed experimentally for key species such as CO, H, OH, and COOH on electrode surfaces. CO adsorption on metal surfaces in particular has been widely treated as a benchmark for theoretical surface science studies(*4, 5*), because CO is an electrocatalytically active intermediate from which viable pathways to C$_1$ and C$_2$ products are formed on Cu or known to be end products on metals such as Ag, Au, or Pt.

Density Functional Theory (DFT) is almost universally used in catalysis modeling today(*6*), typically at the level of a generalized gradient approximation (GGA) due to its lower computational cost and reasonable accuracy for molecular chemistry, but has yielded mixed success for modeling surfaces important to heterogeneous catalysis. A particular Achilles heel has been the inability to predict adsorption energies and adsorption-site preferences on electrocatalytically relevant metals, in which numerous theoretical studies have reported the drawbacks of the GGA functionals on predicting the physical and chemical properties of metal surfaces(*7-10*). Often the most cited reasons are due to the inexact exchange–correlation term(*11*) including lack of non–local correlation which is necessary to describe dispersion interactions(*12*), accounting of zero point energies(*13*), and missing *a posteriori* thermodynamic corrections(*14, 15*). In response, a number of GGA functionals have been developed to better reproduce adsorption energies, such as the RPBE GGA functional which fulfills the Lieb-Oxford criterion by construction.(*7*)

An alternative is to investigate other classes of DFT functionals. While a number of new hybrid DFT functionals that incorporate exact-exchange and range-separation are available(*16-19*), they are not viable given their computational expense for the long-term goal of a complete description of the solid-liquid interface. However, one rung higher from the GGA on Jacob's ladder are the meta-GGA functionals that incur ~4 times the expense, and are possible contenders for a complete DFT model for electrocatalysis. Just recently a benchmark study including RPBE and the meta-GGA functionals SCAN(*20-22*), RTPSS(*23*), and B97M-rV(*24, 25*) models, reported on their ability to reproduce experimental surface relaxation properties, and CO adsorption energies and site preferences, on the M(111) where M = Pt, Cu, Ag, Au metal surfaces.(*26*) While RPBE performed well for the benchmark bulk and surface relaxation properties, the first principles SCAN(*20, 21*) and semi-empirical RTPSS(*23*) and B97M-rV(*24, 25*) meta-GGA functionals yielded mixed results, displaying under-relaxed surface layer displacements and/or strong overbinding of CO on Pt and Cu. Disappointingly, all DFT functionals considered did not predict the CO adsorption site preference – the *atop* site for all metals observed at low CO adsorbate coverage – instead predicting stronger binding to multi-coordinated metal sites. The correct binding site is a necessary prerequisite to more accurately predict the inherent mechanism of a catalytic reaction, and this has been an ongoing problem with all DFT calculations to date. But a further indicator for problems ahead for the RPBE functional is its inability to describe something as basic as the energetics for the water dimer for which it performs very poorly.(*26*) By contrast, the B97M–rV(*24, 25*) functional gives good performance not only on the water dimer, but also an excellent description of bulk water(*27, 28*), revealing its potential feasibility for better theoretical predictions at the solid-liquid interface.

But as is standard in nearly all electrocatalysis computational work, DFT functionals are evaluated at 0 K, while all electrochemical surfaces, adsorbants, and reactions are experimentally produced under finite temperature conditions. Although thermodynamic corrections in the harmonic regime have been used, they can't account for the genuine thermal (and pressure) fluctuations that are manifest in all catalytic experiments.(*29, 30*) Hu et al studied CO adsorption/desorption on Pt(111) using PBE functional within the framework of molecular dynamics,(*31*) and emphasized the importance of the free energy to solve "the CO puzzle" for predicting the atop site.(*9, 32*) Their findings raise the possibility that statistical fluctuations driven by thermal motion might fundamentally change the structural configuration of the metal and thereby the energetic stability and site preference of the CO bound state(*33*), and molecular fluctuations of the liquid/solid interface will undoubtedly influence the chemical pathways taken by electrochemical catalytic reactions. To address this important aspect of electrocatalysis modeling, we have used *ab initio* molecular dynamics (AIMD)(*34*) in the NVT and NPT ensembles to better and more directly simulate the natural statistical fluctuations of an electrocatalytic surface under ambient conditions.

Herein, we consider the entropic thermodynamic forces and how they manifest in the experimental bulk and surface relaxation properties, and CO adsorption energies and site preferences on the M(111) surfaces where M = Pt, Cu, Ag, Au using the GGA functional RPBE(*26*) and the meta-GGA functional B97M–rV(*24, 25*). Our results show that both RPBE and B97M-rV functionals exhibit greater expansion of the top layers of the bare M(111) surface, which in turn modifies the adsorption energy and site preference of CO on all four metals. Both functionals show greatly improved binding energies for the CO binding to the Cu(111) and Pt(111) metal surfaces when evaluated at 300 K, with stronger preference for CO binding to the under-coordinated *atop* site for both metals at finite temperature, although only B97M-rV predicts the trend correctly for the remaining multi-coordinated sites. Thermalization exposes the importance of dispersion interactions, with only the B97M-rV functional predicting weak binding energetics in quantitative agreement with experiment for the Ag(111) and Au(111) metals evaluated here, but are seen to be a mixture of chemisorbed and physisorbed CO species. Overall a more realistic accounting of statistical fluctuations of a thermalized ensemble is shown to improve DFT agreement with experiment, but that functionals with dispersion offer a more balanced performance, setting the stage for future studies involving more realistic and robust models of the solid-liquid interface of many electrocatalytic reactions.

**RESULTS**

The model for the bulk metal and bare M(111) surfaces and adsorbate-surface sites are provided in Figure 1. Six metal layers are employed in total to describe the surface layer and bulk character of the metals, with each layer containing 36 (6 x 6) metal atoms and then imposing periodic boundary conditions in the x-y plane with no constraints used to fix layers. All *ab initio* molecular dynamics (AIMD) simulations are performed in the NVT ensemble, except for the Au(111) and Ag(111) surfaces at the RPBE level of theory which were calculated in the NPT ensemble as their bulk systems were found to be unstable by imposing the known lattice in the NVT ensemble.

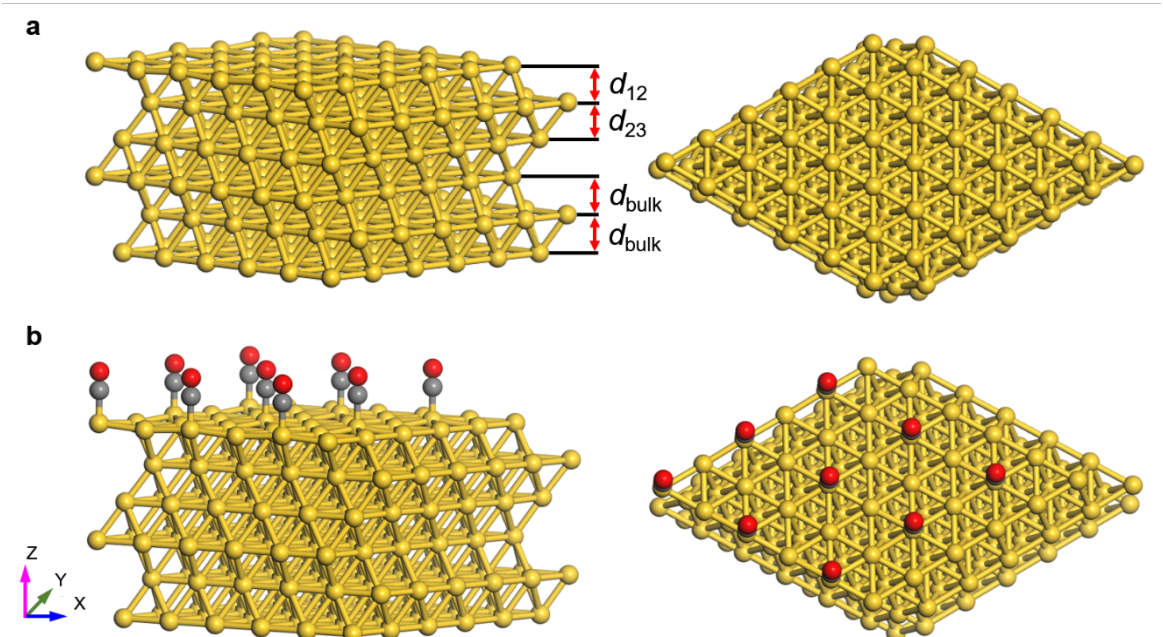

**Figure 1. Bulk metal and metal surface models used to describe structural relaxations and CO binding preferences.** (a) The 6 x 6 model for the bare M(111) surface, with interlayer distances $d_{12}$ and $d_{23}$; the $d_{bulk}$ value is defined as the averaged distance over the last two interlayer distances. (b) Initial condition of nine adsorbed CO molecules on the surface, from two different perspectives corresponding to a low coverage result of 0.25; only the *atop* site configuration is shown here, but a similar set up was used for the initial condition of the other three multi-coordinated sites (bridge, fcc, and hcp shown in Figure S1) for Pt(111) and Cu(111).

Adsorption properties are averaged across nine CO molecules placed onto the surface to construct a low ~25% coverage of available metal sites, also shown in Figure 1, for which the experiments at ambient conditions indicate that the *atop* site is preferred for all metals examined here. All AIMD simulations reported are comprised of 2 ps trajectories, with statistics for observables collected over the last 1.5 ps. Further details are provided in Methods and in the Supplementary Information.

**Interlayer relaxation of metal surfaces.**

We carried out static calculations at 0 K with the RPBE and B97M-rV functionals for the metal models displayed in Figure 1 to determine interlayer distances $d_{12}$, $d_{13}$, and $d_{23}$, values and calculated layer-layer percentage relaxations with respect to $d_{bulk}$ values for each metal at each condition. Overall, the 0 K results generated with the RPBE functional appear to agree well with the experimental data trends for Pt(111) and Cu(111), while also showing acceptable expansion trends for Au(111), but yielding poor agreement for surface relaxations for the Ag(111) metal (Table S1 and Figure 2). In contrast the interlayer distances for the B97M-rV functional at 0 K are more compressed on average across the Pt(111), Cu(111), and Au(111) metals with respect to experiment, and surface relaxation trends are also poorly reproduced for the Ag(111) metal. This result is the same for different plane wave cutoff values (Table S2) and different cell sizes and different numbers of frozen layers to better enforce the bulk lattice constant of the meta-GGA DFT model (Table S3). The strong contraction of interlayer relaxation for B97M-rV relative to RPBE at 0 K is attributable to the inclusion of dispersion term in the functional. To illustrate that it is not an isolated issue with the meta-GGA, we also evaluated the interlayer relaxations comparing RevPBE with and without the D3 van der Waals correction, in which we also see a significant compaction trend with inclusion of dispersion (Table S4).

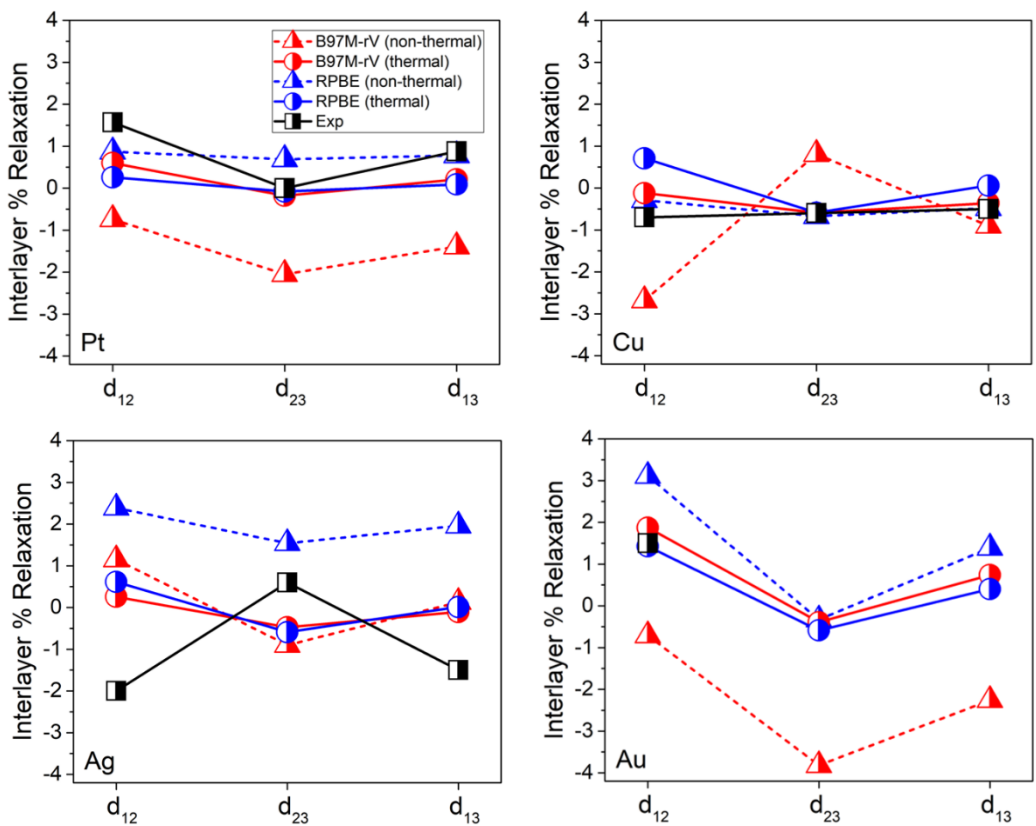

**Figure 2. Calculated interlayer relaxation (%) of M(111) (M = Pt, Cu, Ag, Au) surfaces.** Experimental results are also depicted here for comparison. The absent error bars are all less than 0.07 Å, relative to the DFT functional dependent layer-layer distances in the bulk. $d_{12}$, $d_{23}$ and $d_{13}$ correspond to $d_{12}/d_{bulk}$, $d_{23}/d_{bulk}$ and $d_{13}/2d_{bulk}$, respectively, consistent with Table 1.

But in order to better compare to the thermodynamic condition of the experiments, the surface and bulk properties evaluated with DFT should be collected as thermal averages of surface strain and bulk metal properties, which are also provided in Figure 2 (and Table S1). We note that previous experimental work has estimated a lattice heating time constant on the 1-10 ps timescale for 20 nm metal films for Cu and Au(*41*), which would suggest that the 2.0 ps AIMD simulation timescales (the first 0.5 ps are discarded as equilibration) are sufficient for our simulated ~1 nm thick metals. For the RPBE functional at 300 K the thermal effects weaken the M...M interactions and induce more flexibility, resulting in a larger $<d_{bulk}>$ that increase absolute error with respect to experiment (Table S1). Previous work has reported that RPBE severely underestimates cohesive energies of the four transition metals examined here(*42*), which likely explains the uniform expansion of the $<d_{bulk}>$ values when thermal energy is added, and for the high distortions we observed in the NVT ensemble by imposing the bulk lattice.

The thermal effect diminishes agreement with experiment for the relative surface relaxations for Pt(111) and Cu(111), although there is some improvement for Ag(111) and Au(111) profiting from the interlayer relaxation due to expanded $<d_{bulk}>$ distances (Figure 2). The ambient temperature results for the B97M–rV functional corrects the 0 K surface relaxation percentages significantly, and trends are more consistent with the experimental data for all M(111) surfaces. For both the GGA and meta-GGA functionals, the fluctuations of the interlayer distances $<d_{13}>$ and $<2d_{bulk}>$ are generally anti-correlated (Figure S2) to maintain the stability of the whole surface, and the improvement to the $<d_{13}>$ relaxation for B97M-rV especially is attributable to its convergence to the same average $<2d_{bulk}>$ value not seen at 0 K as shown in Table S1.

**Metal surfaces with CO adsorbates**

Because the correct metal surface strain trends are attributable to the finely balanced occupation of metal *d* states for adsorbate–surface interactions, one would conclude from Table 1 and Figure 2 that the RPBE GGA at either temperature, and the B97M-rV meta-GGA evaluated at 300 K, might be able to predict the adsorption energies and adsorbed-site preference trends for the CO intermediate(*43*), especially for the Cu(111) and Pt(111) surfaces. The popular RPBE functional in particular has provided reasonably reliable results for chemisorption energies at 0 K of relevant catalytic intermediates such as CO previously, although never the low-coordination site preference.

For Cu(111), experimental adsorption energies are available for the *atop*, *hpc*, *fcc*, and *bridging* sites, and CO site preference occurs in the order *atop > fcc > hcp > bridge*(*44, 45*), whereas experimental adsorption energies for all other M(111) metals report only the *atop* site binding energy, as that site is thought to be strongly preferred with respect to the multi-coordinated sites.(*46-48*) We note that previous work for CO on Cu found that timescales for phonon coupling is on the ~ 1 ps timescale(*49*), and thus our AIMD simulation timescales are sufficient for establishing at least CO binding trends on all metals examined here. Thus we emphasize a few general observations as well as a number of surprises that emerge from the comparison of the DFT functionals under the different thermodynamic conditions for CO adsorption energetics and metal site preferences on the M(111) surfaces.

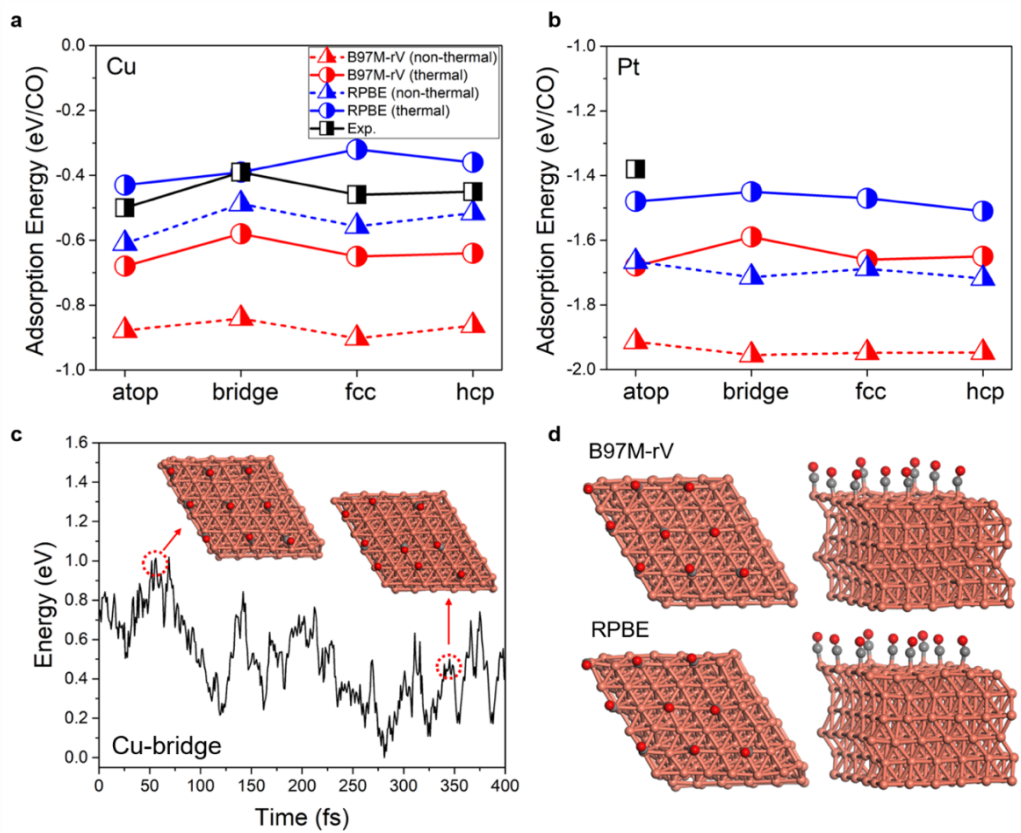

**Figure 3. Thermal effects on the adsorption energies for 25% CO coverage on Cu(111) and Pt(111) surfaces.** Calculated and experimentally observed adsorption energies for CO on the *atop*, *bridge*, *fcc*, and *hcp* sites at 0 and 300 K for **(a)** Cu(111) and **(b)** Pt(111) (see Table S4 for numerical values). **(c)** Binding energy vs. time for B97M-rV, illustrating CO adsorbed on Cu(111) surface at the bridge site for the first 100 fs, but moving to the *atop* position due to the strong energetic preference for CO binding at this site. **(d)** Representative geometries from two different perspectives of the CO adsorbed on Cu(111) at the *atop*-site for both B97M-rV and RPBE, showing the deviations from linearity of the M-C=O angle at 300 K. Absent error bars for the calculations are all less than 0.02 eV/CO. Experimental references for CO binding to Pt(111) and Cu(111) are found in compilations reported by Wellendorff and co–workers(*48*) and Abild–Pedersen and Andersson(*13*). While these CO–M measurements may vary in their starting temperatures that is just experimental convenience to get CO on the surface as it clearly assumed throughout the literature that Ea is a constant over a wide range of temperatures.

Figure 3 summarizes the DFT results using the RPBE and B97M-rV functionals for CO binding energy at 0 K and CO thermodynamic binding stability at 300 K, at 25% coverage for the more strongly binding Cu(111) and Pt(111) metal surfaces; numerical results are also reported in Table S5. Both DFT functionals improve their agreement with experiment for the CO chemisorption energies, in which thermalization corrects for the overbinding at 0 K (Figure 3a and 3b), and in which both prefer the *atop* site as well. However, with thermalization the RPBE functional now slightly underbinds CO to Cu(111) on average, and does not predict the experimental chemisorption energy trend for the other sites. By contrast the meta-GGA functional at 300 K predicts the relative binding energy trend with coordination site consistent with experiment, and correctly ranks the *bridge* site as least stable for Cu(111) as is evident in the dynamics (Figure 3c). The quantitative results can be quite sensitive to the pseudopotential as we have shown previously(*26, 50*), and the plane wave cutoff plays a role too (Figure S3), but the qualitative effect of thermal fluctuations is to create slightly larger M−C−O angles and longer M−C distances as the main indicator of a weakened bonding interaction between the metal at the *atop* site and the adsorbed CO (Figure 3d and Table S5).

For Pt(111) we used a small core pseudopotential for the DFT functionals which we have shown previously results in small errors when benchmarked on a single metal-carbonyl system using a high quality hybrid DFT functional.(*26, 50*) In this case the CO binding energy at the *atop* site of the metal with RPBE using AIMD are in good numerical agreement with experiment, but there is a stronger preference for the CO to bind at the *hcp* site at 0 K and 300 K. Although the CO binding energy for Pt(111) is still overbound by ~0.2 eV/CO using B97M-rV at 300 K, it prefers the *atop* site in agreement with experiment. On the basis of radial distribution function $g_{M-C}(r)$ generated, the integration under the first peak indicates that there is 100% chemisorbed CO at any temperature for both DFT functionals (Figure S4) with a CO coverage of 0.25.

The same calculations were performed for the CO adsorption energy on the *atop* site of the Ag(111) and Au(111) metal surfaces (Figure 4). For the RPBE functional at either temperature, CO is predicted to be more stable in the gas phase for the two weaker binding metal surfaces, given their positive adsorption energies (Table S4). The first $g_{M-C}(r)$ peak gradually shrinks in the simulations over time, indicating that the number of initially bound CO molecules on Au(111) and Ag(111) surfaces is decreasing, and for which we observe the near complete dissociation of CO from Ag(111) on the 2 ps timescale of the AIMD simulation, which is indeed impractical within a short simulation time.

The observation that CO effectively does not bind to the weak metal surfaces using the RPBE functional, with or without thermalization, arises in part from the lack of a dispersion interaction term(*7*), unlike the B97M-rV functional. In fact the B97M–rV functional exhibits binding energetics that agree with experiment on the *atop* site at any temperature, with better agreement found for Ag(111) than for Au(111). More interestingly, the thermalized weak binding surfaces reveal a mixture of chemisorbed and physisorbed CO molecules for the B97M-rV functional, with ~6 CO (~64%) and ~7 CO (~80%) molecules chemisorbed on Ag(111) and Au(111) surface, respectively, with the remaining CO's located up to ~4.0 Å from the adsorption site, which would be the evidence for physisorption. But given the good agreement with experiment for CO adsorption energy on Au(111) and Ag(111) surfaces, the B97M-rV result would propose that experiments might be measuring a mixture of chemisorbed and physisorbed CO species. Thermal fluctuations create much larger M−C−O angles and longer M−C distances for the chemisorbed species at the *atop* site on the Au(111) and Ag(111) surfaces which would diminish the binding strength further compared to the stronger binding Pt(111) and Cu(111) metals (Table S6).

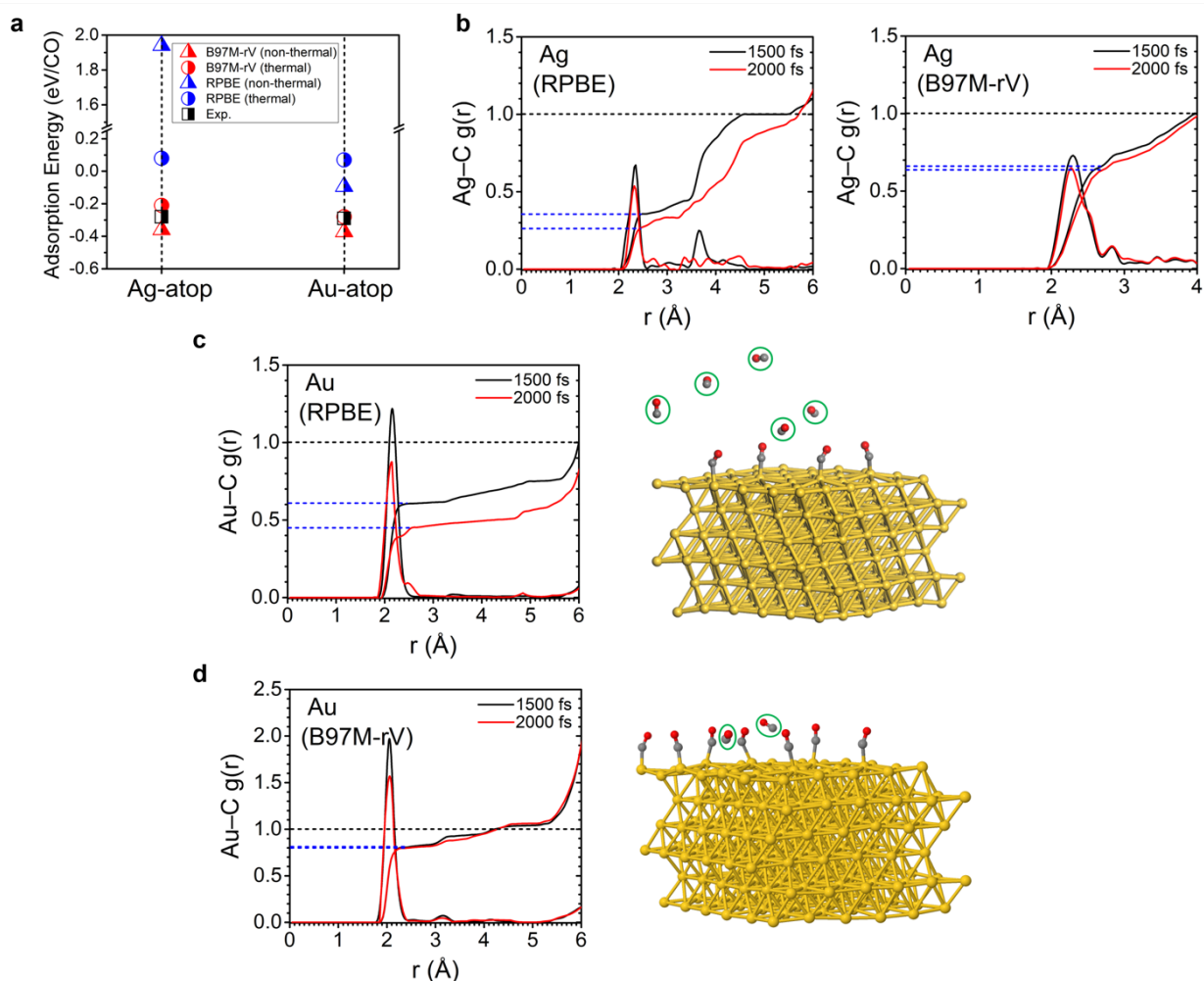

**Figure 4. Thermal effects on the adsorption energies for 25% CO coverage on Ag(111) and Au(111) surfaces.** **(a)** Calculated and experimentally observed adsorption energies for CO on the *atop* site at 0 K and 300 K for RPBE and B97M–rV. Experimental energies for CO binding: Ag(111)(*51*) and for Au(111) as an average over two reported values(*52, 53*). **(b)** Radial distribution function (RDF) between the Ag metal and carbon of CO, $g_{Ag-C}(r)$, for RPBE and B97M–rV after 1.5 ps (black) and 2.0 ps (red), and the corresponding integration to assess the running coordination number across the 9 *atop* sites on the surface. The blue dash lines correspond to the chemical adsorption and the region between blue and black dash lines represent the physical adsorption; beyond this the RDFs are probing second nearest neighbors. **(c)** RPBE and **(d)** B97M-rV results for Au(111), showing $g_{Au-C}(r)$ and representative snapshots to show the mixture of chemisorbed, physisorbed, and/or desorbed CO molecules (circled in green) at the last 2.0 ps time point of the AIMD simulation; snapshots for Ag(111) are shown in Figure S5. The statistical data is collected after 500 fs's pre-equilibration for the 2 ps trajectories.

## DISCUSSION

In this work we have eschewed post-calculation (harmonic) thermodynamic corrections in favor of *ab initio* molecular dynamics that directly accounts for thermal fluctuations of the M(111) surfaces

(M= Ag, Au, Cu, and Pt) and their CO binding motifs. We have compared two different DFT functionals, the GGA RPBE and the meta-GGA B97M-rV, which were chosen for several reasons. First and foremost they are computationally affordable when one looks ahead to the next goal of atomistic modeling of the solid-liquid interface. Second the RPBE functional(*7*) has become a popular choice for surface science studies and electrocatalysis applications(*42*), whereas meta-GGAs such as the B97M-rV functional(*24, 25, 27*) are still undergoing evaluation across a range of applications.(*23, 26, 50, 54*) Finally, we believe that several general conclusions can be drawn in regards the entropic effects for DFT by considering these two functionals specifically.

Previous work has reported that the overall performance of RPBE for lattice parameters, cohesive energies, and surface energies of the four transition metals examined here is worse than that of the parent PBE functional(*42*), in spite of its better performance for describing chemisorption energies for molecules like CO. This would suggest that RPBE at 0 K has a reliance on the severe underestimation of cohesive metal energies(*42*), to create reasonable metal configurations to enable good CO binding trends across a range of metals. But we find that all of the metals systematically expand when RPBE is simulated at 300 K, which is manifest in some underbinding and inconsistent site preferences for CO binding on Cu(111), and complete CO desorption on the Ag(111) and Au(111) surfaces even given the short length of our AIMD simulations, a result consistent with its lack of any dispersion model.

The metal surface contractions evaluated with the B97M-rV functional evaluated at 0 K are greater than RPBE and experiment, and thus thermal energy flow systematically expands the metal surfaces too, but now in such a way to describe more correctly all interlayer distances and surface relaxation properties of the four metals. Consistent with these findings, the B97M-rV functional systematically prefers the *atop* adsorption site for CO for the Cu(111) and Pt(111) surfaces, reproduces the relative binding energy trends for the remaining multi-coordinated sites for Cu(111), and CO is found to weakly bind to the *atop* adsorption site for Ag(111) and Au(111) surfaces with energetics in remarkable agreement with the experimental observations. Although the AIMD simulations used here are quite short, we expect the following qualitative results to hold with B97M-rV, namely, that while the Pt(111) and Cu(111) surfaces show complete chemisorption of CO, the experiments may actually be reporting an average of chemisorbed and a small fraction of physisorbed species on the Ag(111) and Au(111) surfaces. Furthermore, because the B97M-rV meta-GGA gives an excellent description of bulk water properties(*27, 28*), and combined with this work better establishes a theoretical foundation for describing the electrolyte properties at the electrocatalytic surface in the future.

Finally, we believe a few general trends learned in this work on RPBE and B97M-rV in particular will hold when evaluating any DFT functional in general. Thermal effects are seen to populate different normal modes of the metal surface, causing a significant change in their surface relaxation profiles. Because these surface strain trends correlate with CO binding motifs, thermalized statistical averaging influences the preferred CO adsorption site and the relative strength of binding to other sites on any given metal surface. Although predicting the CO absorption preference on the *atop* site of electrocatalytic metals has eluded DFT previously, we suspect that DFT functionals which are more accurate for intermolecular interactions, when properly evaluated at ambient conditions, will on average determine this site preference correctly more often than not.

In summary, the continued quest to find reliable theory to understand mechanisms for CO2RR (and other coupled reactions such as the oxygen evolution reaction) has traditionally focused on better quantum mechanics but less so on the statistical mechanics that is required to match the laboratory conditions for electrocatalysis. In this work we have shown that separation of these theoretical frameworks ignores the true nature of statistical fluctuations on top of what may or may not be a reasonable potential energy surface with a given level of quantum mechanics. That is, the relative performance of any DFT functional at 0 K for describing interlayer relaxations and CO adsorption energies at 300 K is not always reliable – and a more meaningful comparison is the ability to describe an experiment at or near the thermodynamic state point at which data is collected. At the same time, a better study on the relative merits of DFT for describing electrocatalysis must go beyond this work, as there are still unresolved issues around the origin of self-interaction errors of non-hybrid functionals(*55*), pseudopotentials(*26*), zero point energy(*13*), basis sets, distance cutoffs, numerical quadrature, and differences in software packages. Nonetheless our results suggest that DFT functionals might be better than we thought for electrocatalysis, or at least not as bad as we feared, upon full consideration of the detailed fluctuations of a complete statistical mechanical ensemble.

## MATERIALS AND METHODS

All calculations were carried out with DFT using the dispersion corrected meta-GGA functional B97M–rV(*24, 25*) and revised GGA RPBE(*7*) method in combination with TZVP basis sets optimized for multigrid integration(*56*) as implemented in the CP2K package(*57*). The norm-conserving pseudopotentials(*58*) were used for describing the interactions between the frozen cores and electrons in valence shells. In all cases, we used periodic boundary conditions with a cutoff value of 400 Ry and relative cutoff of 60 Ry. All the slabs were repeated periodically with a 15 Å vacuum layer between the images in the direction of the surface normal.

### Static 0 K calculations

Based on previous work we established a bulk model within the GPW formalism, by replicating the unit cell four times along the three cell axes (264 atoms in total).(*26*) Using the parameters obtained from bulk calculations, the slab model was built with a different number of surface layers for testing. Results presented in the main text are for the case of a 6 × 6 × 6 supercell with three bottom layers frozen. For the CO adsorbed systems at a coverage of ~0.25, nine CO molecules were put onto the surface at equal spacing. During the geometry optimizations, the converged criteria were $3 \times 10^{-3}$ bohr for atomic displacements and $4.5 \times 10^{-4}$ hartree/bohr for the forces.

**AIMD simulations at 300 K**

We investigated thermal effects by using AIMD simulations, with a slab model containing six layers with 6 × 6 atoms per layer; no layers were fixed in the simulations. All molecular dynamics simulations with the B97M–rV functional were performed by sampling in the canonical (NVT) ensemble employing Nose−Hoover thermostats(*59, 60*), with a time step of 1 fs and a temperature of 300 K. The same NVT protocol was used for the RPBE functional for Pt(111) and Cu(111), but the NVT simulations led to large distortions of the structure for Ag(111) and Au(111) metals, because that RPBE is a kind of revised GGA by including some unrecognized fitting parameters which is proved to perform well at 0K. However, due to the lack of dispersion term, the metal-metal interaction will be weakened drastically at elevated temperature in the framework of RPBE. In particular, NVT ensemble is sensitive to density, which could cause large structural distortion at RPBE level as it is not able to hold the effective interatomic interaction. It will be improved by using isobaric-isothermal ensemble (NPT) ensemble because an external pressure force is applied to effectually overcome the unexpected expansion. We also examined the ensemble effect of RPBE functional in Pt(111) surface relaxation, as described in the footnote of Table 1. The difference is almost negligible within these two ensembles which made us trustingly switched to NPT ensemble for the weakly bound surfaces Ag(111) and Au(111). Statistical averages were collected over 2 ps of production after 0.5 ps equilibration, where convergence was measured by the settling of the energy fluctuation to ~0.25 kcal mol$^{-1}$ per atom. During the simulation, we found the high flexibility for CO molecules moving to different sites. To overcome this issue, we used the equilibrated structure and add CO adsorbates onto different sites of the surface according to the geometries obtained in the static calculations, then ran an additional 1.5 ps starting from 15 independent equilibrated trajectories of the surface calculations, important for the multi-coordinated sites to address the issue of delocalized chemisorption states.(*31*) The time-averaged distances and adsorption energies are generated from all the trajectories during equilibration. Herein, the time-averaged adsorption energy is calculated as:

$$\langle \Delta E_{ads} \rangle_{300K} = [\langle E(M–9CO) \rangle_{300K} – 9\langle E(CO) \rangle_{300K} – \langle E(M) \rangle_{300K}]/9$$

To quantify the error of the statistical values, we consider the standard deviation based on the distance and energy fluctuations:

$$\sigma = \sqrt{\frac{1}{N}\sum_i^N (\overline{X_i} - \langle \overline{X_i} \rangle)^2}$$

The time-averaged $\overline{X_i}$ value is collected on each 100 fs.

**ACKNOWLEDGMENTS.** This work is supported by the U.S. Department of Energy, Office of Science, Office of Advanced Scientific Computing Research, Scientific Discovery through Advanced Computing (SciDAC) program. This work used the resources of the National Energy Research Scientific Computing Center, a DOE Office of Science User Facility supported by the Office of Science of the U.S. Department of Energy under Contract No. DE-AC02-05CH11231.